\documentstyle[titlepage,12pt]{article}
\def\half{\frac{1}{2}}

\def\be{\begin{equation}}

\def\ee{\end{equation}}
\def\rq#1{(\ref{eq#1})}
\def\to{\rightarrow}
\def\ad{\dot{a}}
\def\sqL{\sqrt{\Lambda}}
\def\sqLp{\sqrt{\Lambda'}}
\def\Rc{{\cal R}}
\def\Rcd{\dot{\cal R}}
\newcount\sectionnumber
\def\sect
{\global\equationnumber=0
\global\advance\sectionnumber by 1
\the\sectionnumber . }
\newcount\equationnumber
\def   \num
{\eqno{\global\advance\equationnumber by 1
\left(\the\sectionnumber .\the\equationnumber \right)}}
\begin{document}

\title{Gravitationally Collapsing Dust in $(2+1)$ Dimensions}

\author{S.F. Ross and R.B. Mann\\
        Department of Physics\\
        University of Waterloo\\
        Waterloo, Ontario\\
        N2L 3G1}

\date{August 12, 1992\\
WATPHYS TH 92/07}
\maketitle

\begin{abstract}
We investigate the circumstances under which gravitationally collapsing
dust can form a black hole in three-dimensional spacetime.
\end{abstract}

\section{Introduction}

General Relativity in $(2+1)$ dimensions has been a source of fascination
for theorists in recent years, primarily because of the potential insights
into quantum gravity that it offers. One has the full structure of the
Einstein equations but with an enormous amount of technical simplification
due to the fewer number of dimensions. One consequence of this simplicity
is that the metric outside of a finite matter source is locally flat
\cite{deser,gidd}, and the mass affects the space-time only globally,
seemingly implying that there are no black hole solutions to $2+1$ gravity.

However, it has recently been pointed out by Ba\~nados {\it et. al}
\cite{3dbh} that if a negative cosmological constant is introduced, so that
the field equation is
\be
G_{\mu\nu} = 8\pi G T_{\mu\nu} - \Lambda g_{\mu\nu},     \label{eq1}
\ee
where $\Lambda > 0$, then there is a solution for the field around a point
source which has an event horizon, {\it i.e.} a black hole.
For zero angular momentum this is
\be
ds^2 = -(\Lambda R^2 -M) dT^2+ \frac{dR^2}{\Lambda R^2 - M}+ R^2d\theta^2 .
\label{eq2}
\ee
Apart from not being asymptotically flat, this solution exhibits many of
the properties of black holes in four dimensions, such as a well-defined
temperature and entropy. It is therefore useful in that it provides another
opportunity to model classical and quantum dynamics of black holes with a
simpler set of field equations.

Here we investigate under what circumstances a disk of pressureless dust,
(the 3D analog of Oppenheimer-Snyder collapse) will collapse to the black
hole solution \rq2. Provided the initial density is
sufficiently large, we show that collapse occurs in finite proper time,
and external observers see the event horizon form in infinite coordinate
time. The other properties of this collapse parallel the results in four
dimensions, as well as recent results in two dimensions \cite{match}.

Collapsing dust solutions in 3D have been studied before \cite{gidd}.
Here we extend this analysis, reproducing the results of ref.
\cite{gidd} when $\Lambda = 0$, and finding new solutions for $\Lambda <
0$. In this latter case we find that collapse to a point source is
possible, subject to certain conditions.

\section{Collapse to the Black Hole}

Consider a disk of collapsing dust surrounded by a vacuum region, with the
metric in the exterior region being given by \rq2. The dust is falling freely,
so we may make it the basis of a comoving coordinate system. We then have a
Robertson-Walker metric on the interior region,
\be
ds^2=-dt^2+ a^2(t) \left( \frac{dr^2}{1-kr^2}+r^2d\theta^2 \right),
\label{eq3}
\ee
where $r$ and $\theta$ are comoving radial and angular coordinates, $t$ is the
proper time of the dust, and $a(t)$ is the scale factor. In these coordinates
$T_{\mu\nu} = \rho u_\mu u_\nu$ is the stress-energy of the dust, where
$\rho(t)$ is the density of the dust and $u_\mu =(1,0,0)$. Conservation of
stress-energy $T^{\mu\nu}_{\ ;\nu}=0$ then implies $\rho a^2 = \rho_0 a_0^2$,
where $\rho_0$ is the initial density of the dust and $a_0$ is the initial
scale factor. The field equations \rq1 become
\be
\ddot{a} = -\Lambda a   \label{eq5}
\ee
and
\be
\Lambda a^2+k - 8\pi G \rho_0 a_0^2+\ad^2 = 0,     \label{eq6}
\ee
where the overdot denotes $d/dt$.

The solution of these equations is
\be
a(t) = a_0 \cos(\sqL t) + \frac{\ad_0}{\sqL} \sin(\sqL t), \label{eq7}
\ee
where
\be
\ad_0 = \sqrt{8\pi G\rho_0 a_0^2 - k -\Lambda a_0^2}     \label{eq8}
\ee
to satisfy the second field equation. As we wish $a(t)$ to be real, we must
require
\be
8\pi G\rho_0 a_0^2 - k -\Lambda a_0^2 \geq 0.    \label{eq9}
\ee
In particular, if we choose the initial conditions $a_0=1, \ad_0=0$, this
relation gives
\be
k = 8\pi G \rho_0 - \Lambda.     \label{eq10}
\ee
Subject to the condition \rq9, this solution always collapses to $a(t_c)=0$ in
the finite proper time
\be
t_c = \frac{1}{\sqL} \tan^{-1} \left[\left(\frac{8\pi G\rho_0 a_0^2 - k
-\Lambda a_0^2}{\Lambda a_0^2} \right)^{-\half}\right]. \label{eq11}
\ee

In the exterior coordinates, the stress-energy vanishes, and the solution is
the black hole \rq2. The dust edge is taken to be at $r=r_0$ in the interior
coordinates, and at $R=\Rc (t)$ in the exterior coordinates. We now wish to
impose conditions to make the dust edge a boundary surface. These are
\cite{israel,chase}
\be
[g_{ij}]=0   \qquad\mbox{and}\qquad   [K_{ij}]=0      \label{eq13}
\ee
where $[\Psi]$ denotes the discontinuity in $\Psi$ across the
edge, $K_{ij}$ is the extrinsic curvature of the dust edge, and the subscripts
$i,j$ refer to the coordinates on the dust edge. The metric on the edge is
\be
ds^2 = -dt^2+\Rc^2(t) d\theta^2, \label{eq14}
\ee
so the condition \rq{13} implies
\be
-(\Lambda \Rc^2 -M) \dot{T}^2+\frac{\Rcd^2}{\Lambda \Rc^2 -M} =-1,\label{eq15}
\ee
where the overdot denotes $d/dt$, which gives
\be
\frac{dT}{dt} = \frac{\sqrt{(\Lambda \Rc^2-M) + \Rcd^2}}{\Lambda \Rc^2 -M},
\label{eq16}
\ee
and $\Rc (t) = r_0a(t)$, that is, the position of the dust edge in the
exterior coordinates is the proper distance from the origin to the dust edge.
Note also that the latter condition implies that the boundary conditions
$a_0=1,\ad_0=0$ represent a ball of dust with initial radius $\Rc_0=r_0$
initially at rest in the exterior coordinates, and we see that $a=0$
corresponds to the collapse of the dust to $\Rc=0$. We now need to compute
\be
K_{ij} = - n_\alpha \frac{\delta e^{\alpha}_{(i)}}{\delta \xi^j}, \label{eq17}
\ee
where $n_\alpha$ is the normal to the edge, $e^{\alpha}_{(i)}$ are the basis
vectors on the edge, and $\xi^i$ are the coordinates on the edge. In the
interior coordinates,
\be
e^{\alpha}_{(0)} = (1,0,0), e^{\alpha}_{(1)} = (0,0,1/r_0 a(t)) \mbox{ and }
n_{\alpha} = \left(0,\frac{a(t)}{\sqrt{1-kr_0^2}},0 \right). \label{seq18}
\ee
A straightforward calculation gives
\be
K_{00} = K_{01} =0,\quad K_{11} = \frac{\sqrt{1-kr_0^2}}{r_0 a(t)}
\label{seq19}
\ee
in the interior coordinates. In the exterior coordinates, we find
\be
e^{\alpha}_{(0)} = (\dot{T},\Rcd,0), e^{\alpha}_{(1)} = (0,0,1/\Rc) \mbox{ and
}
n_{\alpha} = (-\Rcd,\dot{T},0). \label{eq18}
\ee
It follows that
\be
K_{00} = -\frac{d}{d\Rc} \sqrt{(\Lambda \Rc^2-M)+\Rcd^2},
\quad K_{01}=0,\quad K_{11}= \frac{\dot{T} (\Lambda \Rc^2-M)}{\Rc}
\label{eq19}
\ee
in the exterior coordinates. We thus find that \rq{13} implies
\be
(\Lambda \Rc^2-M)+\Rcd^2 = 1-kr_0^2, \label{eq20}
\ee
that is,
\be
M = (\Lambda a^2+ k +\ad^2)r_0^2-1=8 \pi G\rho_0 a_0^2r_0^2-1,  \label{eq21}
\ee
where \rq6 has been used\footnote{Note that this identification of
the parameter $M$ agrees with that used in \cite{3dbh}.}.

Thus, the requirement that the dust edge be a boundary
surface fixes $M$ in terms of the initial density $\rho_0$. Note that
collapse to a black hole occurs only for $\rho_0$ sufficiently large
(as $M$ must be positive), analogous to the $(1+1)$-dimensional case
\cite{match,Arnold}. For $\rho_0<\frac{1}{8\pi G a_0^2 r_0^2}$, the endpoint
of collapse is a naked conical singularity in anti-deSitter space.

So long as $\rho_0>\frac{1}{8\pi G a_0^2 r_0^2}$,
an event horizon will form around the collapsing dust, at
$R_h = \sqrt{M/\Lambda}$.
The comoving time $t_h$ at which the event horizon and the dust
edge coincide is found by substituting $r a(t_h) = R_h$, which gives
\be
t_h = \frac{2}{\sqL}\tan^{-1}\left(\frac{\ad_0 r_0 + (\ad_0^2 r_0^2+ a_0^2
r_0^2
\Lambda -M)^{1/2}}{\sqL r_0 a_0 + \sqrt{M}}\right), \label{eq23}
\ee
and is clearly finite. This is not the case for the coordinate time at
which an external observer will observe this formation. A light signal
emitted from the surface at time  $T$ obeys the null condition
\be
\frac{dR}{dT} = \Lambda R^2 -M \label{eq24}
\ee
and arrives at a point $\tilde{R}$ at time
\be
\tilde{T} = T + \int_{r_0a(t)}^{\tilde{R}} \frac{dT}{dR} dR
 = T - \frac{1}{\sqrt{M\Lambda}} \tanh^{-1} \left(\sqrt{\frac{\Lambda}{M}}
 R \right)_{r_0a(t)}^{\tilde{R}} \label{eq25}
\ee
and thus $\tilde{T} \to \infty$ as $r_0 a(t) \to \sqrt{M/\Lambda}$, so the
collapse to the event horizon appears to take an infinite amount of time, and
the collapse to $\Rc=0$ is unobservable from outside.

The comoving time interval $dt$ between emissions of wave crests
is equal to the natural wavelength $\lambda$ that would be emitted in the
absence of gravitation, and the interval $d\tilde{T}$ between arrivals of wave
crests is the observed wavelength $\tilde{\lambda}$. Thus the red shift of
light
from the dust edge is
\be
z=\frac{d\tilde{T}}{dt} -1=\frac{1}{\sqrt{1-kr_0^2}+r_0\ad(t)}-1,\label{eq26}
\ee
and $r_0\ad(t_h) = -\sqrt{1-kr_0^2}$, so $z \to \infty$ as $t \to t_h$.
Thus the collapsing fluid will fade from sight, as the red shift of light
from its surface diverges. These properties entirely
parallel those of Oppenheimer-Snyder collapse in four dimensions.

\section{Other Collapse Scenarios}

If we now consider the case where the cosmological constant is set to zero,
we see that the event horizon is no longer present, and collapse occurs to
a naked source. This is the scenario investigated in \cite{gidd}; writing
the interior metric in the form \rq3, the solution of the field equations is
\be
a(t)  = a_0 + \ad_0 t, \mbox{ where } \ad_0  = \pm\sqrt{8 \pi G\rho_0 a_0^2-k}.
\label{eq27}
\ee
where we must require
\be
8 \pi G \rho_0 a_0^2 \geq k \label{eq28}
\ee
for $a$ to be real (equation (104) of \cite{gidd}). This condition is
equivalent to the requirement that
$\alpha^2 > 0$ in \cite{gidd}. As pointed out in \cite{gidd}, pressureless
dust need not collapse; only
if $\ad_0$ is negative will collapse occur in time
$t_c= (8\pi G \rho_0 - k/a_0^2)^{-1/2}$.
The exterior metric \rq3 for $\Lambda=0$
has the appropriate signature only if $M$ is negative, in
which case we can convert it into the form in \cite{gidd},
\be
ds^2 = -dT'^2+dR'^2+c R'^2 d\theta^2,       \label{eq31}
\ee
by taking $T'^2 = -MT^2, R'^2 = -R^2/M$ and $c = -M$. If we choose the dust
edge to be at $r=r_0$, $R'= \Rc(t)$, we may apply the matching conditions
(\ref{eq13}) to the metrics (\ref{eq27},\ref{eq31}). The metric on
the dust edge is \rq{14}, so \rq{13} gives $\sqrt{c} \Rc = r_0 a(t)$ and
\be
\frac{dT'}{dt} = \sqrt{1+\frac{r_0^2\ad_0^2}{c}}.        \label{eq32}
\ee
Imposing the condition that the dust edge move at subluminal
speed\footnote{For
$\Lambda > 0$, we need not impose this condition, as the dust edge
asymptotically comes to rest at the horizon from the point of view of an
external observer.}
($| \frac{dR'}{dT'}| < 1$) yields
\be
\frac{r_0^2 \ad_0^2}{1-k r_0^2} < 1.     \label{eq37}
\ee
which is easily seen to be equivalent to equation (103) of \cite{gidd}
upon using the relationship
\be
r = \frac{\sin(\sqrt{k} \chi_0)}{\sqrt{k}}      \label{eq38}
\ee
between the two interior metrics.  The extrinsic curvature is calculated
as before,  and \rq{13} implies
\be
c = 1-8\pi G \rho_0 a_0^2r_0^2.       \label{eq33}
\ee
The condition \rq{37} implies $c < 1-k r_0^2 $ in our coordinates, which
is the same as \rq{28}.

Finally, we consider the case $\Lambda < 0$ in the field equation \rq1. This
corresponds to a positive cosmological constant in the usual terminology, and
we write \rq1 as
\be
G_{\mu\nu} = 8\pi G T_{\mu\nu} + \Lambda' g_{\mu\nu},     \label{eq41}
\ee
$\Lambda' > 0$. In the interior metric \rq3, the solution now becomes
\be
a(t) = a_0 \cosh(\sqLp t) + \ad_0 \sinh(\sqLp t),  \label{eq42}
\ee
where
\be
\ad_0 = -\sqrt{8\pi G \rho_0 a_0^2-k+a_0^2 \Lambda'}     \label{eq43}
\ee
to satisfy the other field equation. We now see that, for $a(t)$ to be real,
we must have
\be
8\pi G \rho_0 a_0^2-k+a_0^2 \Lambda' \geq 0. \label{eq44}
\ee
We also see that we must have $\ad_0 < -a_0 \sqLp$ to achieve collapse to a
point. This is not surprising, as the positive cosmological constant is
causing an overall expansion of the space-time. When collapse is possible,
$a(t_c)=0$ at
\be
t_c = \frac{1}{\sqLp} \tanh^{-1} \left[\left(\frac{8\pi G\rho_0 a_0^2 - k
+\Lambda' a_0^2}{\Lambda' a_0^2} \right)^{-\half}\right]. \label{eq45}
\ee

The exterior metric in this case is
\be
ds^2 = -(-\Lambda' R^2 - M)dT^2+ \frac{dR^2}{-\Lambda' R^2 -M} + R^2 d\theta^2,
\label{eq46}
\ee
where $M$ must be negative in order to get the right signature (yielding an
exterior deSitter space metric), so we will
write $c=-M$ from here on. There is a cosmological event
horizon at $R_h = \sqrt{c/\Lambda'}$. Taking the dust edge to be at $r=r_0$,
$R=\Rc(t) < R_h$, and applying the conditions (\ref{eq13}) to the
metrics (\ref{eq42},\ref{eq46}), we find $\Rc(t) = r_0 a(t)$,
\be
\frac{dT}{dt} = \frac{\sqrt{(\Lambda' \Rc^2+c) + \Rcd^2}}{\Lambda' \Rc^2+c},
\label{eq47}
\ee
and
\be
c = 1-(-\Lambda' a^2+ k +\ad^2)r_0^2=1-8 \pi G\rho_0 a_0^2r_0^2  \label{eq48}
\ee
as before. From \rq{44}, we see that $c < 1- (-\Lambda' a_0^2+k)r_0^2$ to
satisfy this condition. The collapse condition $\ad_0 < -a_0 \sqLp$ implies
$c < 1-kr_0^2$, which is more restrictive.

As the collapse in this case is to a naked conical singularity, we should
again require
\be
\left| \frac{dR'}{dT'} \right| < 1,      \label{eq49}
\ee
that is, the velocity of the dust edge in the exterior coordinates of
\cite{gidd} should be less than the speed of light. Thus, we require that
\be
\left| \frac{r_0 \ad(1-kr_0^2-r_0^2\ad^2)}{1-kr_0^2} \right| < c. \label{eq50}
\ee
As $\ad(t)$ is a monotonically increasing function of $t$ in $(0,t_c)$, we
may treat the left-hand side as a function of $\ad$, and we then need only
impose the condition at the point in the interval $(\ad_0,\ad_c)$  where it
is a maximum.

The situation then divides into three cases, depending on the relative size
of $c$.  If $c < 2/3 (1-kr_0^2)$, the maximum is at $\ad_c$, and we must
impose
\be
c > (1-kr_0^2) - (1-kr_0^2)^2,   \label{eq51}
\ee
which requires as a corollary $1-kr_0^2 > 1/3$. If $c> 2/3(1-kr_0^2)$, but
$c< 2/3(1-kr_0^2)+\Lambda' a_0^2 r_0^2$, then the maximum is at an
intermediate point, and we must impose
\be
c> \frac{2}{3\sqrt{3}} \sqrt{1-kr_0^2}, \label{eq52}
\ee
which requires as a corollary $1-k r_0^2 > 4/27$. If $c> 2/3(1-kr_0^2)+
\Lambda'
a_0^2r_0^2$, and $\Lambda' a_0^2 r_0^2 < 1/3(1-kr_0^2)$, then the maximum is
at $\ad_0$, and we must require
\be
c> \left| r_0 \ad_0 \left( 1-\frac{r_0^2\ad_0^2}{1-kr_0^2} \right)\right|.
\label{eq53}
\ee

\section{Conclusions}

Pressureless dust in $(2+1)$ dimensions can undergo a variety of collapse
scenarios, depending upon the relative signs and magnitudes of its initial
density, initial collapse speed and the cosmological constant ($-\Lambda$).
The stationary black hole solution found in \cite{3dbh} arises naturally
from gravitational collapse of pressureless dust for a negative
cosmological constant. Its properties are similar to those of the
higher-dimensional Oppenheimer-Snyder case in that collapse to a point
singularity occurs in finite proper time, and the event horizon forms in
infinite
coordinate time, with an infinite redshift. Requiring that the dust edge be
a boundary surface gives a relation between the parameter $M$ in the
exterior coordinates and the initial density of the dust.

For $\Lambda\le 0$ collapse, if it occurs at all, is to a naked conical
singularity. The solution for $\Lambda =0$ in our coordinate system  is
equivalent to that found in \cite{gidd}.  The condition that the dust edge
move subluminally in the exterior coordinates may be imposed as an
additional condition, although this does not provide any new information.
The results for the case $\Lambda < 0$ follow the same pattern, although
here collapse  requires that the initial velocity be great enough to
overcome the overall  expansion of the spacetime caused by the cosmological
constant.  Imposing the condition that the velocity of the dust edge be
less then the speed of light yields a somewhat complicated relationship
between this quantity and the initial density. Collapse to a naked conical
singularity is also possible for $\Lambda > 0$ if the initial density is
sufficiently small.

\section{Acknowledgments}
This work was supported by the Natural Sciences and Engineering Research
Council of Canada.

\end{document}